# The Scandinavian Style: Nordic values in HCI

**Rebekah Rousi**
School of Marketing &
Communication
University of Vaasa
rebekah.rousi@uwasa.fi

**ABSTRACT**
During the 1950s Scandinavian Design caught international attention with its minimalism, simplicity, functionalism and sophistication. Several factors rested at its heart: functionality, democracy and affordability. Aesthetic styles connected to international minimalist, modernist and functionalist movements, which were also symbolically connected to education, political and social movements, and the Nordic welfare model. Studies have shown how social and political values from this period connect with Nordic interaction design from the past three decades. How these are represented in contemporary interaction design discourse, and design form and expression is a perspective under represented. This paper presents the results of a three-tiered content analysis of the proceedings of NordiCHI years 2000-2014: categorization of titles according to emphasis; content analysis of Scandinavian value constructs overall; and thematic connection of the results to conference theme and site. Results are then discussed with reflection on form and process in the Nordic interaction design industry.

**ACM Classification Keywords**
H.1.2 User/Machine Systems: human factors; human information processing.

**INTRODUCTION**
Scandinavian Design has its roots firmly embedded in centuries' worth of handicraft and trade traditions, as well as more recent international movements such as the European Arts and Crafts Movement [14, 17]. In addition, other influential factors informing the styling of Scandinavian Design have stemmed from developments in European aesthetic movements such as National Romanticism [40]. Thus, design and nationalism have been intrinsically connected for over two centuries.

Furthermore, thanks to several high profile international exhibitions titled *Design in Scandinavia* (1954-1957), Scandinavian Design has been globally recognized for its quality, innovative use of materials, democratization of the design process, affordability, and minimalistic, yet harmonious and functional aesthetic dimensions [1, 21]. Scandinavian or Nordic Design has spawned from centuries of craftsmanship and development on the practical, everyday (folk) as well as social-political levels, and embodies ideologies most importantly from the Scandinavian (Nordic) welfare models [34]. The foundational idea behind Scandinavian Design and aesthetics, was that it was in the nations' best interests that in order to encourage national prosperity, greater attention needed to be placed on individual wellbeing – or wellbeing of the folk [4, 24, 34]. That is, in order to have a higher functioning, productive and advanced nation of people, attention needed to be placed on increasing the living standards of these people and providing: intellectual wellbeing through education for all; physical wellbeing through adequate medical facilities and services; and emotional wellbeing through the design or creation of everyday aesthetics [2, 34]. Moreover, in Scandinavian Design major focus has traditionally been placed on the home - the home as the incubator or nurturer of intellectual and cultural talent.

Artists such as the Swedish Carl and Karin Larsson [39] were some of the driving forces behind what is known as the Swedish "home for the people" (*folkhemmet*), which can be interpreted as a social-political movement or manifesto towards providing equal access to domestic beauty and aesthetic pleasantness in the everyday lives of the people, regardless of class [11, 35]. The idea was that with attention placed towards outlooks and expression, the way in which people experienced and approached the world would be taken to a higher cognitive and intellectual level. That is, the design of things should not only be functional, but they should also be beautiful, allowing people to benefit from the aesthetics through their understanding of the greater dimensions in life and the world, which only culture can afford.

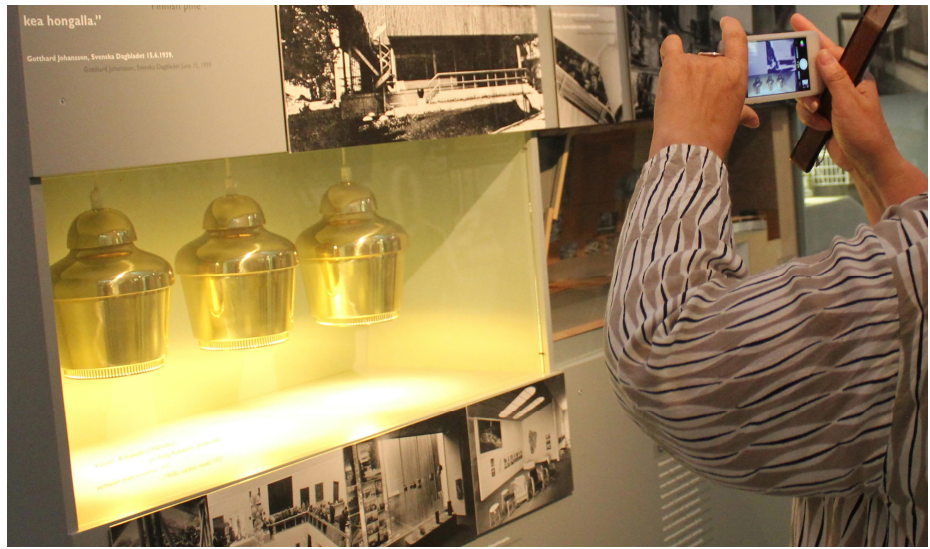

**Figure 1. Display of architecture and Artek Alvar Aalto A330 ceiling lamps (1954)**

When taking a look at this phenomenon through cultural theory, philosophy and sociology, we may refer to Pierre Bourdieu [6] and his modes of capital. Through these we can see that economic, social and cultural capital, are closely intertwined with one another. In other words, if the level of cultural capital is improved, ergo people are familiarized with and able to appreciate the arts (visual, music, theatre etc.), literature and design for example, then subsequently, their social capital is also improved through, e.g., greater intellect (familiarity and knowledge with cultural discourse), which also enables them to interact with and relate to a broader population of people, and increases their awareness of social-cultural phenomena, and subsequently the ability for self-reflexivity [6]. From the perspective of cognition also, this serves to bring perception and experiential processes more often into the realm of higher order cognitive processing, than simply that of lower order [9]. And finally, with greater awareness, aptitude and ability to navigate in the social, cultural and political landscapes, also comes greater potential to benefit economically.

Throughout the Nordic region, particularly in Sweden and Finland, this awareness of the intrinsic relationship between the *Bourdieu style* capitals, productivity and industrialness, has served to spur both internal-external notions of national identity and culture, and has also formed the rationale behind modern industry [23]. On a rudely simplistic level, it can be said that, culture and aesthetic structuring and experience through the arts and design, education to support this in production, interpretation and appreciation, and industry are intimately linked [18]. The air of, the form of, the cultural-political propaganda around [23, 26] design in the Nordic countries, saw everyday objects become trophies of modern living. Sophisticated forms and presentation of lighting (see figure 1), and fine-grained attention to human factors such as ergonomics (see figure 2), made the Nordic region stand out, not simply in terms of its remoteness and proximity to nature (this is often talked of in terms of the Mystification and Otherness of the North (see e.g. [27]), but also producers of superior technology and craftsmanship [40].

It is against these roots, that contemporary Nordic technology design and innovation is rested. How this reputation is manifested and expressed in the realm of human computer interaction (process, models and scholarship) and technology form is the topic of investigation. This paper discusses previous attempts to characterize a Nordic, or Scandinavian, style of human computer interaction, paying attention to key themes attributed to being Scandinavian by nature. It also presents a content analysis of the papers and titles of NordiCHI conferences from the year 2000 to present. Finally, it takes a brief look at some of the trends and expressions seen in high technology, and human computer interaction (e.g., user experience) companies today.

**CAPTURING SCANDINAVIAN DESIGN IN HUMAN COMPUTER INTERACTION**

Rising in popularity during the 1950s, Scandinavian Design, or the *Scandinavian Style* caught international attention with its minimalism, simplicity and functionalism [23]. A quintessential vehicle for its exposure was the *Design in Scandinavia Exhibition* (1954-57). This exhibition was not only a platform for Scandinavia to showcase its best in terms of design prowess - skills, forms and materials - but was also an important vehicle through which Finland in particular could align itself, and establish its foundations, in the discourse of Western design, production and industry [23]. Thus, design in the Nordic region has always been and still is both the leverage point of industry, as well as culminated symbol of what the Nordic model has to stand for: collaboration, cooperation, skills, craftsmanship and quality [3, 4, 10, 30].

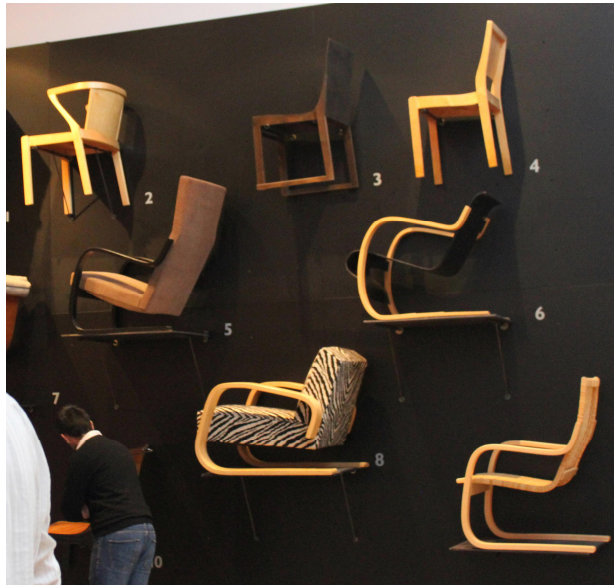

**Figure 2. Display of Artek Alvar Aalto chairs in form pressed wood (Alvar Aalto Museum)**

Several factors rested at the heart of traditional Scandinavian Design, and these were: beauty (wellbeing through aesthetics and intellect), functionality, democracy and affordability [33, 40]. Common materials and processes used were: plastic, anodized/enameled aluminum, pressed steel, woven textiles and form-pressed wood. Thus, the styling of products belonging to the Scandinavian Design paradigm is an important component as they embody and subsequently function as signifiers of Nordic values [36].

The aesthetic movements the paradigm clung onto were modernism and functionalism (The International Style), through the language of minimalism [16]. Scandinavian, or more aptly put, Nordic Design, is globally recognized through forms such as Alvar Aalto's architecture and furniture, Tapio Wirkkala's designs, Arne Jacobsen, and even IKEA's design expression. But, how does this translate to the design language of information technology, and particularly interaction design? Categorizations of Scandinavian Design have been made in human computer interaction, particularly from the perspective of user-centred design and design processes, but how this translates to the physical or final form of the product remains unclear.

This paper presents a content analysis of previous NordiCHI papers from 2000 to present, in an attempt to define: what is discussed in the Nordic realm of HCI and how this reflects a specifically Nordic approach to HCI design; and whether or not there are any distinct characteristics of the outcomes of the designs – is there a typically Scandinavian-Nordic HCI, and how does this connect to established Scandinavian Design traditions?

Susanne Bødker, Pelle Ehn, Dan Sjägren and Yngve Sundblad's NordiCHI (2000) paper "Co-operative Design – perspectives on 20 years with 'The Scandinavian IT design model'" marked two decades of conscious scholarly and design contributions in relation to a Scandinavian or Nordic IT model. The paper describes a project that began in 1981 and ran until 1986 called Utopia (Training, Technology and Product in Quality of Work Perspective - an acronym from Swedish, Danish and Norwegian), which was a larger version of several smaller projects undertaken during the 1970s [3, 4]. The series of projects, and particularly this Utopia project, were dedicated to increasing the standards and skill levels of people working in the field of graphic workstation technology.

The projects did not simply focus on the technical components of increasing professionals' capabilities to produce high quality graphics, rather, researchers focused on a number of factors that they saw as contributing to an ecosystem which fostered effective, high quality design culture. These included investigating prerequisites in terms of social and technical factors, limitations and obstacles [bodker]. The Utopia experience had four key areas: where workers craft technology - design based on work and organizational requirements; setting the stage for design in action - mockups and prototypes; playing the language game for design and use - 'Communities of Practice'; and bringing design to software - bringing design thinking and practice into software development [4].

Paradigmatically, it is interesting to observe Pelle Ehn's [10] "Scandinavian Design: on participation and skill", in which Ehn characterizes the Scandinavian design of computer-based systems around industrial democracy, interdisciplinary action-based research, and most importantly collaborative, cooperative and participatory design processes. Perhaps not surprisingly, considering how Ehn's article was written during the early 1990s, emphasis in the Scandinavian design model is on democratizing design for the work place. This is based on a history of socialtechnical solutions for work-oriented design stemming largely from 1970s Sweden. Moreover, the participatory model presented [3, 4, 10] emphasizes the role of establishing and developing language games, through the utilization of tools such as scenarios, mockups and prototypes. Thus, against the background of earlier Scandinavian or Nordic design traditions, we can see that manifestation of the Nordic ethos - the linguistic or syntactic expression of Nordicness - had moved from materials, towards a more dynamic, and ephemeral discourse of social interaction in the interaction design process. Yet, the foundational values behind the two traditions (e.g. traditional Scandinavian Design and Nordic interaction design) can be seen as shared. These values include: democracy and equal access for all; political freedom and participation; mental and physical wellbeing; and collaboration and community [7, 31].

Moreover, clear similarities can be seen between the Nordic Welfare model and today's human computer interaction design discussions relating to: health, education, employment, social integration, housing, economic security, culture, politics and recreation [29]. For this reason, it seems pertinent to analyze the topics and discourse of NordiCHI conferences from its inception in 2000 to the most recent conference in 2014. The idea is to gain an idea of how these values are represented and manifested within the scholarship and development of interaction design in the Nordic region - either by Nordic residents and citizens or others wishing to participate in the discourse. It also is important to gain insight into how this manifests within the stylistic language of contemporary interactive product design - to see how companies are connecting new technologies with more traditional qualities and expressions that the Nordic region is so widely known for.

**METHOD**

Data was extracted, consisting of all the paper and demo titles, from the proceeding lists of NordiCHI at the Interaction Design Foundation website [25] for the years 2000-2010, and the ACM Digital Library for the years 2012 and 2014. From the data extraction 784 titles were collected. All of the papers were sorted into groups according to the main themes and categories represented in the titles. There were theme overlaps in terms of interaction context, but categorization occurred in relation to the main emphasis of the arguments and approach presented in the title, e.g., "a participatory approach to developing virtual reality applications", would have been sorted into the category for participatory and collaborative design approaches. The process was iterated for each year (total of eight times), and categories were renegotiated each time iteration occurred.

In order to probe the data from another angle, the next step of the procedure was to perform content analysis [13, 19], in order to understand how key values attributed to Nordic and Scandinavian design - interaction and traditional - are represented as a whole on the conference level, across conferences. These results are represented in relation to the conference theme, year and even location.

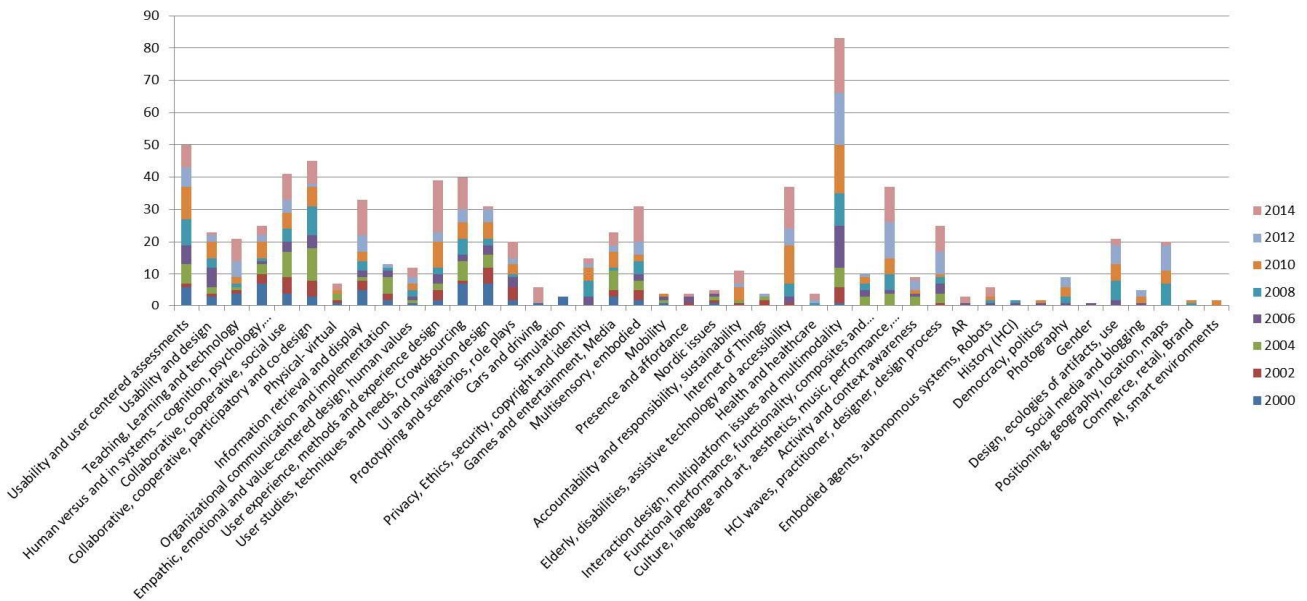

**Figure 3: Categories deriving from the title sorting from NordiCHI 2000-2014**

*Data analysis*
After a final iteration (the ninth) of the titles and categories in Microsoft Excel, then calculating the frequencies of the amount of papers featured in each category, per year, the categories were revised in terms of their relationship to the above mentioned Nordic (Scandinavian) design values. Following this, content analysis [13, 19] of the titles was once again performed, but this time in terms of construct frequencies in general, to determine how the values are represented in academic and practice human computer interaction design discourse in the Nordic region. The design value - construct search - was conducted on the terms: participation, collective, collaborative, democracy, skill, health, education and quality. This was a mixture of constructs found from key readings of Nordic design (interaction design and traditional Scandinavian design) [4, 10, 16, 40]. Furthermore, particular attention was given to the Nordic nature, i.e., directly addressing Nordicness or one of the Nordic countries in the title and subject of the paper.

The structure of the content analysis of the overall titles and topics of the conferences was based on the key values of both Scandinavian and Nordic design (interaction and traditional) as well as Nordic welfare model components of: participation, collaboration, cooperative, collective, democracy, skill, education, culture, health, minimalist, functionalist and affordable. Furthermore, to accommodate for different forms of these constructs, modifications of the words were also searched and coded, e.g., minimal, minimalism, functional, functionalism, collaborative etc.

**RESULTS**
As a result of the iterative categorization process 42 categories were deliberated. These categories demonstrated the diversity of themes and topics that have been presented at the conferences of the years. The categories are not unproblematic, but due to the scope of the diversity represented some generalizations have been made. The categories are: 1) usability and user-centered testing and evaluation methods; 2) usability and design; 3) teaching, learning and technology; 4) human versus/and in systems - cognition, psychology and physiology; 5) collaborative, cooperative, social use; 6) collaborative, cooperative, participatory and co-design; 7) physical - virtual; 8) information retrieval and display; 9) organizational communication and implementation; 10) empathic, emotional, human-values and value-centered; 11) user experience, methods and experience design; 12) user studies, techniques (including crowdsourcing) and needs; 13) UI and navigation design; 14) prototyping, scenarios and role plays; 15) driving and cars; 16) simulation; 17) privacy, security, ethics and legal (copyright); 18) games, entertainment and media; 19) multisensory and embodied; 20) mobility; 21) presence and affordance; 22) Nordic issues; 23) accountability, responsibility and sustainability; 24) Internet of Things; 25) elderly, disabilities, accessibility and assistive technology; 26) health and healthcare; 27) Interaction design, multiplatform issues and multimodality; 28) functional performance, functionality and composites; 29) culture, language, art, aesthetics, music, performance and literature; 30) activity and context awareness; 31) HCI waves, practitioners, designers and design processes; 32) augmented reality; 33) embodied agents, autonomous systems and robots; 34) HCI history; 35) democracy and politics; 36) photography and videos; 37) gender; 38) design, ecologies of artifacts and use; 39) social media and blogging; 40) positioning, geography, location and maps; 41) commerce, retail and brand; and 42) AI and smart environments. As seen in figure 1, some topics dominated more than others. These are discussed in the following.

*Interaction design, multiplatform issues and multimodality*
Understandably, interaction design and its associated components of multiplatform issues and multimodality, was the most popular category across conferences, in terms of emphasis. Overall, across conferences 10.6% of the papers (83) have been directly about developments and implications in interaction design. This has steadily increased during the years 2010-2014 (15, 16 and 17 papers respectively), and was also a popular category in 2006 (13 papers).

*Usability and user-centered assessments*
Usability and user-centered assessments have also been popular themes, whereby 6.4% of the papers (50) have been dedicated to these issues. Papers featuring these peeked during 2008 (8 papers) and 2010 (10 papers), yet have remained at a steady 6-7 papers per year for all the other conferences except 2002 (1 paper).

*Collaborative, cooperative, participatory and co-design*
Titles emphasizing collaborative, cooperative, participatory, and co-design (design perspective) have represented 5.2% (41) of the titles collected. There have been between four to five titles focusing on these issues per conference, part from during the years 2004 and 2014 (each with 8 titles respectively).

*Collaborative, cooperative and social use*
Emphasis placed on collaborative, cooperative and social use (user perspective) was made in 5.7% of the titles (45). The rate to which these have been presented at the conferences has varied quite substantially. But, it can be seen that these issues were particularly popular during the years 2004 (10), 2008 (9) and 2014 (7).

*User experience, methods and experience design*
User experience as well as methods for its measurement and experience design were also popular issues. These represented 5% of the titles (39). For all years except 2010 and 2014, there were between two to three titles emphasizing these issues. However, in 2010 (8) and 2014 (16) saw the peeks of this topic.

*User studies, techniques (including crowdsourcing) and needs*
Titles emphasizing user studies, techniques and user needs represented 5.1% of all the titles (40). The most popular conferences for these topics were 2000 (7) and 2014 (10). The conferences during which these were less represented were 2002 (1) and 2006 (2), and otherwise there have been approximately 4 to 6 papers on these issues per conference.

*Elderly, disabilities, assistive technology and accessibility*
Titles focusing on issues relating to the elderly, disabilities, assistive technology and accessibility accounted for 4.7% of the titles (37). The years in which these were most popular were 2010 (12) and 2014 (13). 2008 (4) and 2010 (5) had some papers around these topics however otherwise these issues have been relatively rarely represented.

*Culture, language, art, aesthetics, music, performance and literature*
Titles focusing on culture, language, aesthetics or the arts (music, performance and literature) have also represented 4.7% of the titles (37). At the past two conferences (2012 and 2014) these issues seem to have gained in popularity (11 titles respectively). There have been a few conferences featuring either 4 or 5 titles emphasizing these (2004, 2008 and 2010), yet at the first two conferences (2000 and 2002) there were no papers emphasizing these - with the exception of the Nordic-related papers (one paper for each of the events in 2000 and 2002).

**Content analysis of Scandinavian/Nordic values**
The content analysis of the overall titles and topics of the conferences was based on the key values of both Scandinavian and Nordic design (interaction and traditional) as well as Nordic welfare model components of: participation, collaboration, cooperative, collective, democracy, skill, education, culture, health, minimalist, functionalist and affordable. Furthermore, to accommodate for different forms of these constructs, modifications of the words were also searched and coded, e.g., minimal, minimalism, functional, functionalism, collaborative etc. This was accomplished by conducting word search in Microsoft Word with the shortest possible denominators for the word forms.

As a result, the five most popular values conveyed in the content of the titles across conferences were: *culture* (and the arts, aesthetics); *education*; *collaboration*; *health*; and *participation*. Overall, culture with its linguistic and artistic associations and word forms was featured across the conferences 33 times. Culture-related themes were most popular in 2012, when the NordiCHI conference theme was "Making sense through design", held in Copenhagen. Education was the next most frequently mentioned construct (with associated terms such as learning and teaching), mentioned 27 times in total. This was mentioned most in 2014 at the Helsinki, "Fun, Fast, Foundational" conference (11 times). Collaboration (collaborative) was mentioned 23 times across conferences with a steady rate of 3-4 times per conference besides 2000, the first NordiCHI held in Stockholm (once), and 2012 (twice) held in Copenhagen. Health related constructs (care, medical etc.) were mentioned 20 times overall - the greatest amount of times in 2014 in Helsinki. Finally, participation (participatory, participative) was mentioned 19 times overall, with the most amount of mentions in 2014.

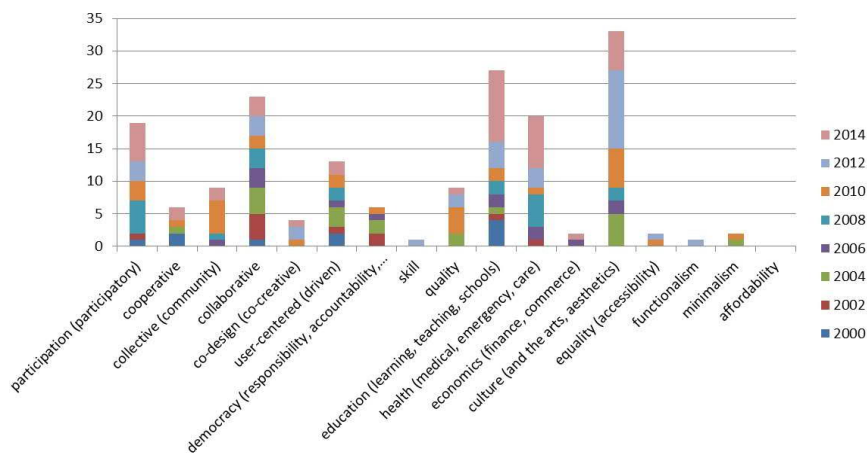

**Figure 4: Representation of Nordic design and welfare values in terminology used at NordiCHI conferences 2000-2014**

Figure 4 shows the distribution of these values across the NordiCHI conferences from 2000 to 2014. Interesting to highlight are the values that have not been so frequently mentioned, or not mentioned at all, such as 'affordability' (not at all), and equality (accessibility) twice - saying this, as noted in the category analysis mentioned above, there were numerous papers presented that focused on people with disabilities and the elderly. In terms of execution, skill was mentioned once, while regarding style functionalism was referred to once, and minimalism twice.

Moreover, reference to specifically Nordic-related issues, and Nordic-reflexivity - in terms of a Nordic (Scandinavian) style of human-computer interaction - have also been under-represented. A Nordic style of human-computer interaction was mentioned specifically at the first NordiCHI in the paper of Bødker et al. [4], and then at three subsequent conferences in direct relation to the state of Sweden's interaction design by Sjöberg and Norlin [38], as well as usability by Gulliksen et al. [22], and in municipality website design by Eliason and Lundberg [12].

**DISCUSSION - FROM WORDS TO PRACTICE**
The focus areas, themes and terms used in the titles are interesting in their power to convey the emphasis placed on particular values within a Nordic strain of human computer interaction. But, what is of even more interest for the purposes of this paper, is how Nordic values in practice, process and form, manifest in today's interaction design environment. For this reason we will briefly observe the products and processes of two contemporary Scandinavian companies: design-people.dk (Denmark) and their Vifa series; and Veryday (Sweden) with their collaborative design emphasis and merging of classic design form with contemporary techniques.

### *Veryday, Sweden*
As the name suggests, Veryday Sweden focuses on designing interactive solutions, products and services to address everyday challenges. Veryday is an interaction design firm that is known as one of the leading global innovation and design consultancies. Here, people are placed at the center of design and service innovation processes, in public image and product outcome. Moreover, key focus points of the company are on healthcare technology design, which emphasis on an empathic approach and superiority, financial services in the sharing economy, public transportation and a mixture of other product and service designs [42].

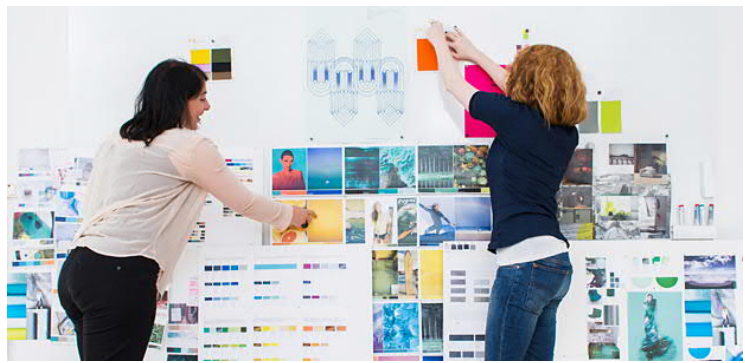

**Figure 5: Team-work in deliberating product design - "Designing exceptional things" (image courtesy of Veryday)**

What is of interest to this paper are the focal points of the company's service offering, as well as their approach and processes. Veryday can be read as the embodiment of Scandinavian design ideals which feature: collaboration, co-design, community, creativity, strategy, quality and skill, with the incorporation of the realities of contemporary global society - economy (finance), sustainability (public transport, sharing and reliability), healthcare and education. Furthermore, an issue that appears to be underrepresented in the papers at previous NordiCHI conferences is the issue of equality - particularly, gender equality. The public image of Veryday strongly emphasizes an equal ratio of active and involved designers and teams (see figure 5).

Moreover, while the company has a strong focus on advanced technologies and interaction design, there process images incorporate numerous materials and methods which link traditional design and creative practices to this new discourse of material conscious information technology design.

### *design-people.dk*
design-people.dk is an interaction design company which encapsulates the spirit of Nordic democracy, with a branded interaction design approach labelled, Female Interaction [8]. Here, the focus is on emancipation, through giving women a voice in product (phone) design, as well as giving her a role in holistic user experience, and re-thinking tech-products from a female perspective. The company is interesting firstly, from this approach to addressing and incorporating gender within the interaction design process, thus, explicitly acknowledging that while in the Nordic countries there has been emphasis on equality and active citizen participation by women, there is still a long way to go before achieving equal representation in the design of everyday things [41]. Thus, education plays a key role in the company's operations, not just from the perspective of users informing the designers in the co-design process, or designers informing the users on how to approach product usage, nor simply in the design of educational software and products, rather, education occurs through discussion and operationalization of these people (female) centric co-design - interactive design - processes. Attention is drawn to the hidden aspects of democracy, the often taken-for-granted equality that does not need any work as it is engrained in the identity and ethos of the Scandinavian countries.

Moreover, not only is the linguistic (language game) [4] and action-based discourse of the company involved in explicating and positively problematizing the two-folded nature of apparent democracy, but also the way in which they treat the product styling and delivery of their high tech products is additionally quite interesting. This company has managed to tap into the styles and sensibilities of traditional Scandinavian design, whereby materialization of the apparently immaterial - interaction design - can be seen as a draw card of the company's portfolio. The Vifa collection - a luxury brand of high tech speaker and sound products - capitalizes on the synergy between technology and materials. The design process stemmed from female participation in the design process, particularly from the starting point of technological products being designed by men, for men, and in an attempt to neutralize the silvers, blacks, greys, harsh edges, abundance of buttons and cheap plastic, textiles are used to integrate the designs into the aesthetics of other everyday objects such as bags (handbags), furniture and wall pieces (see figure 6).

Interestingly, the home is another factor connecting the company, their designs, processes and publicity to the Scandinavian Design paradigm. Through projects such as those with Danfoss - an engineering company offering a broad range of solutions - work has been placed on de-alienating the engineering complexity of for example, home tech solutions (e.g. climate control in the home). With their Danfoss project, the resulting solution focused minimalism (of form, colour and interaction complexity), connection with nature (snowflakes or frost like patterns indicating the status of the climate control system), and trust through sophistication of the design and the system.

Similarly to the case of Veryday, the integration of material culture and styling [36], with the interactive products, helps convey the relevance and integrity of their approach to interaction design. What is more, interaction designs are taken away from the internet - although Apps (financial) and mobile-remote interfaces for e.g., the climate control system etc. are included in their design scope, still what is important is the physical being and interaction of the people involved in the design process and consumption. The design needs to *hit home*, whether that be at work, leisure or during the everyday tasks of running a family.

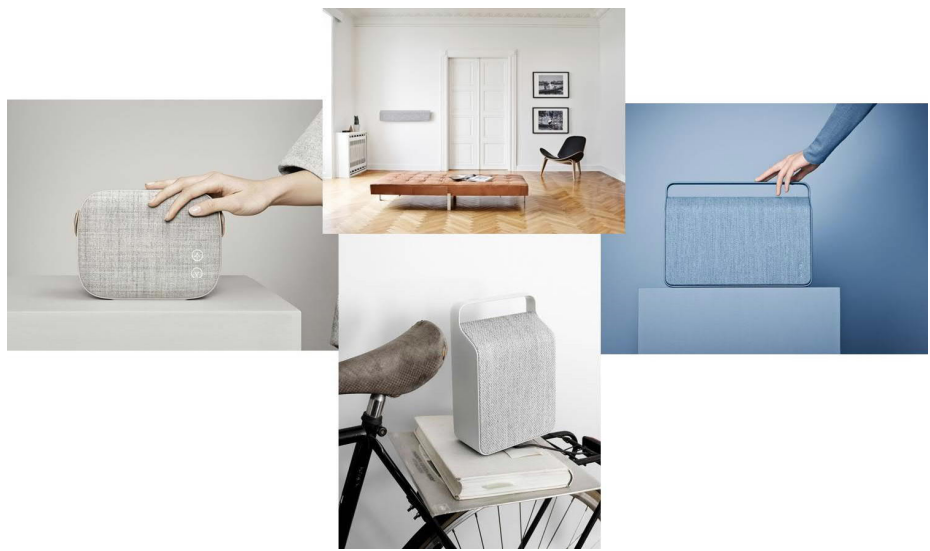

**Figure 6: Vifa, a new luxury brand inspired by women: Vifa Helsinki (*left*); Vifa Stockholm (*top-center*); Vifa Oslo (*bottom-center*); and Vifa Copenhagen (*right*) - (images courtesy of design-people.dk)**

**CONCLUSION**

The purpose of this paper was to reflect on, and gauge the status of Scandinavian Design discourse in contemporary interaction design scholarship and practice. The paper began by characterizing the nature of what is internationally known as Scandinavian, or Nordic, Design, which led into discussion on how this topic has been approached in the work of scholars in the field of human-computer interaction [4, 10]. What was noticed was that particular values of Nordic Design and welfare discourse have been focused on, particularly from the perspectives of: collaboration, cooperation, co-design and participation, as well as skills, and the integration of education with advanced knowledge in information technology development. What was missing was reference to the *styling* and its role in conveying, or communicating the values of Scandinavian Design traditions [36] which are based upon collaboration, equity, productivity, ethics, skills and wellbeing for all [4, 10].

Thus, in an effort to determine how these values are represented and embodied in contemporary Nordic interaction design, a three tiered content analysis was performed on the basis of paper and demo titles from all the NordiCHI conferences (years 2000 to 2014). This comprised: 1) categorization of title emphasis (subject); 2) word search-based content analysis; and 3) connection to conference theme and site. Through data extraction and content analysis some of the foundational writings of the Scandinavian style of human-computer interaction and interaction design were found [4]. Yet, it was noticed that no much explicit attention had been placed on addressing Nordic issues in interaction design. Rather, Nordic values emerged in the themes and wording of the titles such as collaborative, participatory, co-design and social use, as well as core application areas such as accessibility (elderly and disabled), healthcare and education. Interestingly, there seemed to be an underrepresentation of gender issues related to interaction design and human-computer interaction in general. The only two papers addressing gender were in regards to memory performance and habitual media use [28] and gender differences in understandings of telepresence collaborative technology [32]. Likewise, democracy, politics, accountability and responsibility were represented in a few paper titles [e.g., 15, 37], but overall these were not popular genres. In regards to formalistic style of the presented cases, it is difficult to grasp the relationship between the actual solutions developed and presented in the papers, and how they fit on the level of Scandinavian Design traditions. In other words, the tangible, material forms of the technologies are not such as explicit or integral component of Nordic interaction design in the NordiCHI discourse.

Instead, what needs to be considered is the nature of an event such as NordiCHI and its emphasis on interaction design as a whole. The interaction design can be seen simply as the technical components required for facilitating human-computer, and human-human mediated interaction. Or, we can view this from a greater perspective in terms of how the technology, and the conferences, bring the world together in Scandinavian fashion to address current topics faced by people concerning the technological conditions of the era.

What is more, is that in order to give a more rounded and balanced perspective of technology under development, breakthrough technology and its discourse, in relation to Scandinavian interaction design for the everyday, two Nordic interaction design companies were discussed: Veryday, and design-people.dk. Through their approaches, processes and finished products, we gain a glimpse of the way in which Scandinavian Design traditions have been translated and evolved into technological design thinking. This paper has served to establish Scandinavian, or Nordic, Design in the field of interaction design as a discourse, with both a value base and a body. While technical components and systems enabling human-computer interaction to exist are inarguably important, it is also essential to understand that people, meet these designs and systems through a *body*. The body, the outlook or style of the designs, connects to a broader discourse of values, beliefs and traditions which not only communicate quality and equality, but also give the designs to a specific identity and leverage - Nordic Interaction Design.